\documentclass[]{aastex631} 

\usepackage{textcomp}  
\usepackage{comment}
\usepackage{subfigure}
\usepackage{float}
\usepackage{booktabs}
\usepackage{fourier}

\newcommand\cc{black}
\usepackage{lineno}

\submitjournal{ApJ}
\DeclareMathAlphabet{\mathcal}{OMS}{cmsy}{m}{n}
\SetMathAlphabet{\mathcal}{bold}{OMS}{cmsy}{b}{n}

\shorttitle{KamLAND \nue - GRB221009A time-coincident search}
\shortauthors{The KamLAND collaboration}

\graphicspath{{./}{figures/}}

\begin{document}
\title{Limits on the low-energy electron antineutrino flux from the brightest GRB of all time}

\correspondingauthor{M. Garcia}
\email{milesg@udel.edu}

\newcommand{\tohoku}{\affiliation{Research Center for Neutrino
    Science, Tohoku University, 980-8578, Japan}}
\newcommand{\ipmu}{\affiliation{Kavli Institute for the Physics and Mathematics of the Universe (WPI), 
    The University of Tokyo Institutes for Advanced Study, 
    The University of Tokyo, Kashiwa, Chiba 277-8583, Japan}}
\newcommand{\osaka}{\affiliation{Graduate School of 
    Science, Osaka University, Toyonaka, Osaka 560-0043, Japan}}
\newcommand{\rcnp}{\affiliation{Research Center for Nuclear Physics, 
    Osaka University, Ibaraki, Osaka 567-0047, Japan}}
\newcommand{\tokushima}{\affiliation{Department of Physics, 
    Tokushima University, Tokushima 770-8506, Japan}}
\newcommand{\tokushimags}{\affiliation{Graduate School of Integrated Arts and Sciences, 
    Tokushima University, Tokushima 770-8502, Japan}}
\newcommand{\lbl}{\affiliation{Nuclear Science Division, Lawrence Berkeley National Laboratory,
    Berkeley, California 94720, USA}}
\newcommand{\hawaii}{\affiliation{Department of Physics and Astronomy,
    University of Hawaii at Manoa, Honolulu, Hawaii 96822, USA}}
\newcommand{\mitech}{\affiliation{Massachusetts Institute of Technology, 
    Cambridge, Massachusetts 02139, USA}}
\newcommand{\ut}{\affiliation{Department of Physics and
    Astronomy, University of Tennessee, Knoxville, Tennessee 37996, USA}}
\newcommand{\tunl}{\affiliation{Triangle Universities Nuclear Laboratory, Durham, 
    North Carolina 27708, USA; \\
    Physics Departments at Duke University, Durham, North Carolina 27708, USA; \\
    North Carolina Central University, Durham, North Carolina 27707, USA; \\
    and The University of North Carolina at Chapel Hill, Chapel Hill, North Carolina 27599, USA}}
\newcommand{\vt}{\affiliation{Center for Neutrino
   Physics, Virginia Polytechnic Institute and State University, Blacksburg,
   Virginia 24061, USA}}
\newcommand{\washington}{\affiliation{Center for Experimental Nuclear Physics and Astrophysics, 
    University of Washington, Seattle, Washington 98195, USA}}
\newcommand{\nikhef}{\affiliation{Nikhef and the University of Amsterdam, 
    Science Park, Amsterdam, the Netherlands}}
\newcommand{\gppu}{\affiliation{Graduate Program on Physics for the Universe, Tohoku University, Sendai 980-8578, Japan}}
\newcommand{\bu}{\affiliation{Boston University, Boston, Massachusetts 02215, USA}}
\newcommand{\chapel}{\affiliation{UNC Physics and Astronomy, 120 E. Cameron Ave., Phillips Hall CB3255, Chapel Hill, NC 27599}}
\newcommand{\obihiro}{\affiliation{Department of Human Science, Obihiro University of Agriculture and Veterinary Medicine, Obihiro, Hokkaido 080-8555, Japan}}
\newcommand{\sandiego}{\affiliation{Hal{\i}c{\i}o\u{g}lu Data Science Institute, Department of Physics, University of California San Diego, La Jolla, California, 92093, USA}}
\newcommand{\delaware}{\affiliation{Department of Physics and Astronomy, University of Delaware, Newark, Delaware 19716, USA}}
\newcommand{\tohokuimr}{\affiliation
    {Institute for Materials Research, 
    Tohoku University, Sendai, 
    Miyagi, 980-8577, Japan}}

\newcommand{\atkansai}{\altaffiliation
{Present address: Faculty of Environmental and Urban Engineering, 
Kansai University, 3-3-35 Yamate, Suita, Osaka, 564-8680, Japan}}
\newcommand{\aticrrnow}{\altaffiliation
    {Present address: Kamioka Observatory, Institute for Cosmic-Ray Research, 
    The University of Tokyo, Hida, Gifu 506-1205, Japan}}
\newcommand{\atrikennow}{\altaffiliation
    {Present address: Center for Advanced Photonics, 
    RIKEN, Wako, Saitama, 351-0198, Japan}}
\newcommand{\atbutsuryonow}{\altaffiliation
    {Present address: Faculty of Health Sciences, 
    Butsuryo College of Osaka, Sakai, Osaka 593-8328, Japan}}

%
%
\author{T.~Araki}\tohoku
\author{S.~Chauhan}\tohoku
\author{K.~Chiba}\tohoku
\author{T.~Eda}\tohoku
\author{M.~Eizuka}\tohoku
\author{Y.~Funahashi}\tohoku
\author{A.~Furuto}\tohoku
\author{A.~Gando}\tohoku
\author{Y.~Gando}\tohoku\obihiro
\author{S.~Goto}\tohoku
\author{T.~Hachiya}\tohoku
\author{K.~Hata}\tohoku
\author{K.~Ichimura}\tohoku
\author{H.~Ikeda}\tohoku
\author{K.~Inoue}\tohoku
\author{K.~Ishidoshiro}\tohoku
\author{Y.~Kamei}\atrikennow\tohoku
\author{N.~Kawada}\tohoku
\author{Y.~Kishimoto}\tohoku
\author{M.~Koga}\tohoku\ipmu
\author{A.~Marthe}\tohoku
\author{Y.~Matsumoto}\tohoku
\author{T.~Mitsui}\atbutsuryonow\tohoku
\author{H.~Miyake}\tohoku\gppu
\author{D.~Morita}\tohoku
\author{R.~Nakajima}\tohoku
\author{K.~Nakamura}\atkansai\tohoku
\author{R.~Nakamura}\tohoku
\author{R.~Nakamura}\tohoku
\author{J.~Nakane}\tohoku
\author{T.~Ono}\tohoku
\author{H.~Ozaki}\tohoku
\author{K.~Saito}\tohoku
\author{T.~Sakai}\tohoku
\author{I.~Shimizu}\tohoku
\author{J.~Shirai}\tohoku
\author{K.~Shiraishi}\tohoku
\author{A.~Suzuki}\tohoku
\author{K.~Tachibana}\tohoku
\author{K.~Tamae}\tohoku
\author{H.~Watanabe}\tohoku
\author{K.~Watanabe}\tohoku

\author{S.~Kurosawa}\tohokuimr
\author{Y.~Urano}\tohokuimr

\author{S.~Yoshida}\osaka

\author{S.~Umehara}\rcnp

\author{K.~Fushimi}\tokushima
\author{K.~Kotera}\tokushimags

\author{B.E.~Berger}\lbl
\author{B.K.~Fujikawa}\ipmu\lbl

\author{J.G.~Learned}\hawaii
\author{J.~Maricic}\hawaii

\author{Z.~Fu}\mitech
\author{S.~Ghosh}\mitech
\author{J.~Smolsky}\mitech
\author{L.A.~Winslow}\mitech

\author{Y.~Efremenko}\ipmu\ut

\author{H.J.~Karwowski}\tunl
\author{D.M.~Markoff}\tunl
\author{W.~Tornow}\ipmu\tunl

\author{S.~Dell'Oro}\vt
\author{T.~O'Donnell}\vt

\author{J.A.~Detwiler}\ipmu\washington
\author{S.~Enomoto}\ipmu\washington

\author{M.P.~Decowski}\nikhef
\author{K.M.~Weerman}\nikhef

\author{C.~Grant}\bu
\author{\"{O}.~Penek}\bu
\author{H.~Song}\bu

\author{A.~Li}\sandiego

\author{S.N.~Axani}\delaware
\author{M.~Garcia}\delaware
\author{M.~Sarfraz}\delaware

\newcommand{\nue}{$\bar{\nu}_e$ }
\newcommand{\ANALYSISSTARTDATE}{Sept. 3rd, 2002 }
\newcommand{\ANALYSISENDDATE}{xxx. xxxrd, 2021 }

\collaboration{99}{The KamLAND Collaboration}\noaffiliation

\date{\today}

\begin{abstract}

The electron antinuetrino flux limits are presented for the brightest gamma-ray burst (GRB) of all time, GRB221009A, over a range of 1.8\,-\,200\,MeV using the Kamioka Liquid Scintillator Anti Neutrino Detector (KamLAND). Using a variety of time windows to search for electron antineutrinos coincident with the GRB, we set an upper limit on the flux under the assumption of various neutrino source spectra. No excess was observed in any time windows ranging from seconds to days around the event trigger time. The limits are compared to the results presented by IceCube.
\end{abstract}



\section{Introduction} \label{sec::intro}


On October 9, 2022, gamma-ray detectors observed an extremely bright gamma-ray burst (GRB) adjacent to the Galactic Plane on the sky. This event, GRB221009A, was first observed by Fermi-GBM at 13:16:59.99 UT ~\citep{GRBFermiGBM}, and was soon followed by observations from SWIFT BAT and XRT ~\citep{GRBSwift}, as well as Fermi-LAT~\citep{GRBFermiLAT}, with further observations from Konus-Wind and INTEGRAL~\citep{GRBIPN}. The GRB was classified as a long GRB (duration more than 2 seconds), and its position was triangulated by the InterPlanetary Network (IPN)~\citep{GRBIPN}. The initial light curve from Fermi-GBM showed a short pulse at the time of the initial trigger ($T_0$), followed by an extremely bright and long pulse starting 221\,s after $T_0$. $T_{90}$, the time for 90\% of the energy in gamma rays to be observed, was measured to be $\sim$327\,s. This was the brightest GRB ever recorded by Fermi-GBM~\citep{GRBFermiGBM}, saturating the detector for large parts of the second pulse. The position of the GRB, using data from SWIFT, was determined to be RA = 19h 13m 03s, Dec = +19d 48' 09". The redshift of the GRB was determined 11.55\,hrs after $T_0$ by X-SHOOTER at the Very Large Telescope to be z = 0.151~\citep{GRBredshift}.

Followup analyses quickly confirmed the extreme energetics of GRB221009A. It was determined to have the highest fluence, peak flux, and equivalent isotropic energy ever recorded for a GRB~\citep{Burns2023}. LHAASO announced the observation of the highest-energy photon from a GRB ever observed in coincidence with this event at more than 18\,TeV, as well as 5000 VHE (E$>$100\,GeV) photons from the event~\citep{LHAASOVHE}, when only hundreds of VHE photons had ever been seen from GRBs previously. The full LHAASO analysis placed constraints on properties of the GRB, such as the jet opening angle and bulk Lorentz factor, that help explain GRB221009A's unique energetics~\citep{LHAASOScience}. Additionally, IceCube performed an analysis looking for neutrinos with energies between 0.5 and 10$^6$\,GeV, but were not able to claim any significant observations~\citep{IceCubeGRB}.

Prior to this analysis, the Kamioka Liquid Scintillator Anti-Neutrino Detector (KamLAND) has performed astrophysical searches looking for neutrinos coincident with GRBs~\citep{KamLANDGRB}, as well as gravitational waves~\citep{KamLANDGW}, solar flares~\citep{KamLANDSolarFlare}, and supernovae~\citep{KamLANDSupernova}, placing the most stringent limits on the neutrino flux from GRBs at low energies (E$<$100\,MeV). In this paper, we perform a time-coincident neutrino search using KamLAND to calculate the electron antineutrino flux from GRB221009A at the lowest energies from 1.8 to 200\,MeV using various time windows surrounding the event, as well as various source emission spectra. 

\section{Neutrino Emission from GRBs}\label{sec:background}

Gamma-ray Bursts are some of the most energetic electromagnetic phenomena in the Universe. Long GRBs result from core-collapse supernovae~\citep{xu2013discovery}, and are expected to release a large amount of energy  ($\mathcal{O}$(10$^{53}$\,erg)) with neutrinos emitted over a large range of energies from MeV to EeV. The prompt $\gamma$-ray emission of GRBs is thought to come from the dissipation of kinetic energy in relativistic fireballs by internal shocks ~\citep{Rees_1994}, as well as magnetic reconnection in the resulting relativistic outflow~\citep{Meszaros1997}. While accelerated electrons produce the gamma rays associated with the GRBs via synchrotron emission, accelerated protons and neutrons in the outflow can produce high energy (TeV - PeV) neutrinos by interacting with photons~\citep{Waxman1997,Becker2005,Murase2006,Hummer2011,Bustamante2015,Biehl2017,Pitik2022,Ai2023,Liu2023,Rudolph2023}. Additionally, GeV neutrinos may be generated by proton and neutron interactions prior to gamma-ray emission~\citep{Murase2013,Bahcall2000,Bartos2013,Murase2022}, and neutrinos with energies up to EeV may be produced by reverse shocks in the afterglow phase of the GRB~\citep{Waxman2000,Murase2007,Thomas2017,Murase2022,Zhang2022}. At low (MeV) energies, neutrinos may be produced by quasi-thermal processes in the core collapse itself~\citep{Kistler2013-zn}, in neutrino-dominated accretion outflows~\citep{Liu2017,Qi2022}, and within the fireball itself~\citep{Halzen1996}. However, it is unclear as to the level each of these mechanisms play in neutrino production, so conservative time windows are often used to account for the large uncertainties in the production mechanisms. 
\newpage


\section{KamLAND detector} \label{sec:detector} 

\begin{figure}[t]
    \centering
    \gridline{
        \fig{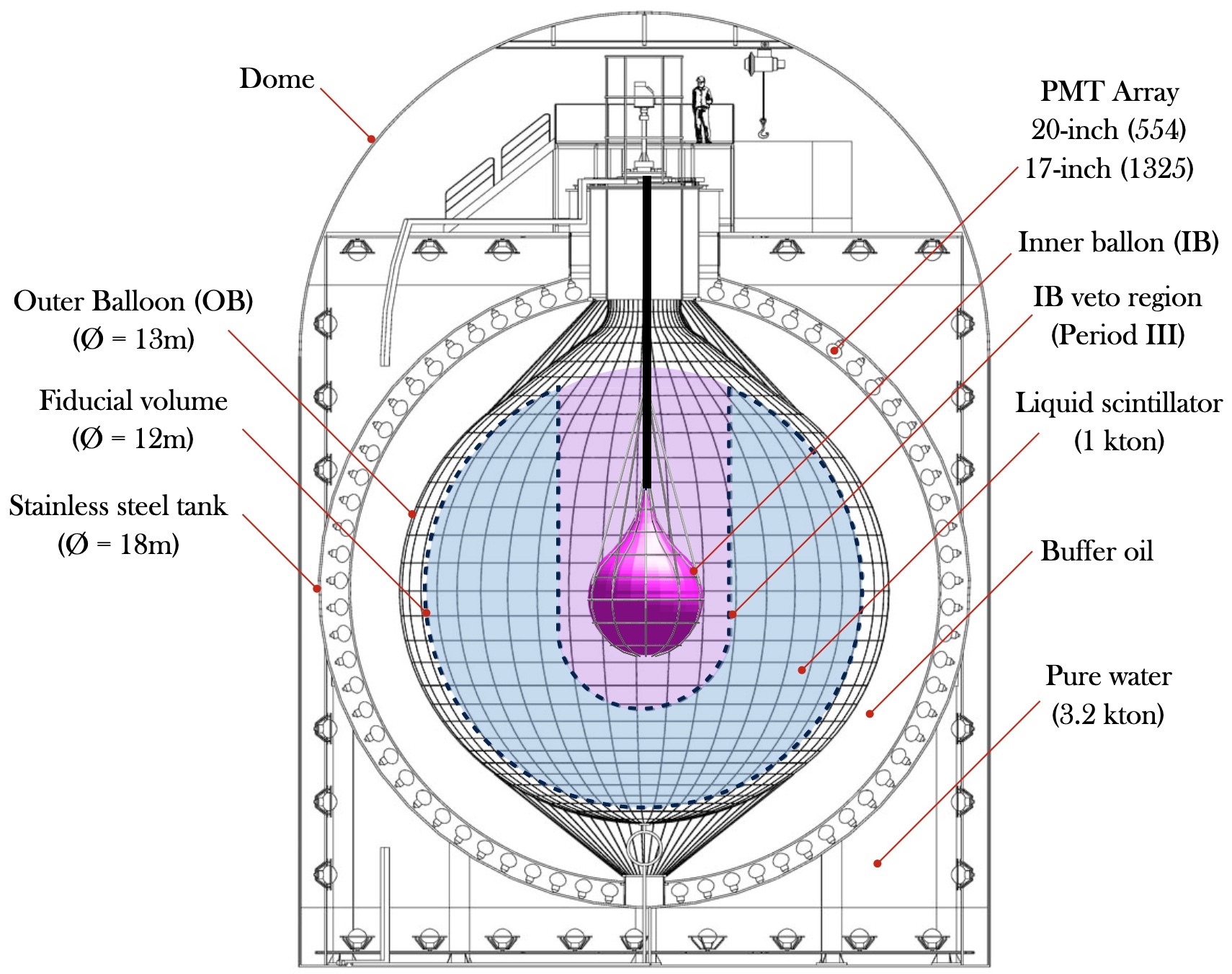}{.55\textwidth}{}
        \fig{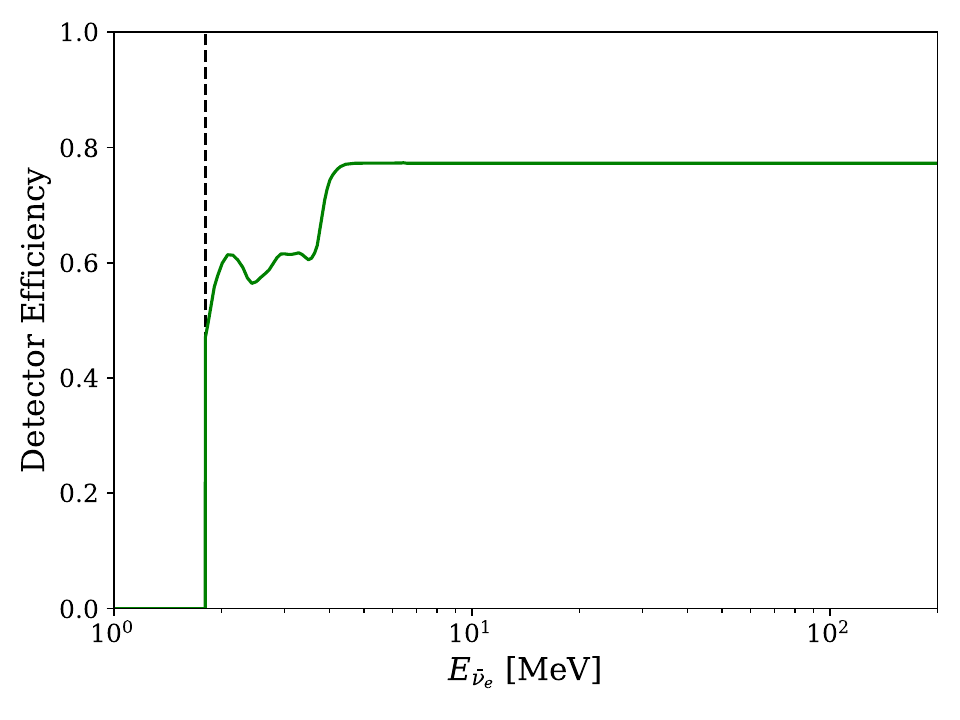}{0.44\textwidth}{}
    }
    \caption{\textit{Left}: A schematic diagram of the KamLAND detector during the KamLAND-Zen 800 experiment. The fiducial volume for \nue detection is highlighted in blue, while the inner balloon cut is shown in purple. The inner detector is azimuthally symmetric. \textit{Right}: The IBD \nue detection efficiencies during the KamLAND-Zen 800 experiment. The structure below $\sim$4\,MeV arises from the likelihood selection. A vertical dotted line is shown at the 1.8\,MeV low-energy IBD threshold.}
\label{fig::detector}
\end{figure}

KamLAND is a 1-kton liquid scintillator detector located 1\,km underneath Mt. Ikenoyama in Gifu, Japan. The detector consists of an outer detector and an inner detector, shown in Fig. \ref{fig::detector} (left). The outer detector holds 3.2\,ktons of water and functions as a water Cherenkov cosmic ray veto for the inner detector. The inner detector itself is an 18\,m-diameter stainless steel tank lined with 554 20-inch and 1325 17-inch PMTs with a smaller, 13\,m-diameter nylon \& EVOH outer balloon of 1\,kton of ultrapure liquid scintillator (80\% dodecane, 20\% pseudocumene) held inside the tank with ropes and buffer oil, which also acts as additional gamma-ray shielding. The innermost 12\,m of the outer balloon is used as the fiducial volume for antineutrino detection. This setup gives reconstructed positional and energy uncertainties of $\sim$12\,cm/$\sqrt{E\,\mathrm{(MeV)}}$ and $\sim$6.4\%/$\sqrt{E\,\mathrm{(MeV)}}$, respectively~\citep{gando2013reactor}. KamLAND was originally built to detect neutrinos from Japan’s nuclear reactors, as well as geoneutrinos~\citep{gando2013reactor,araki2005experimental,abe2008precision}, and started data collection in 2002 towards this end. Starting in October 2011, a 3.08\,m-diameter inner balloon was installed inside the outer balloon, and filled with 326\,kg of xenon-loaded liquid scintillator for the KamLAND-Zen 400 neutrinoless double beta decay experiment~\citep{gando2016search}. From April 2018 to January 2024, a 3.8\,m-diameter balloon was filled with 745\,kg of xenon and inserted in place of the 3.08\,m balloon for the KamLAND-Zen 800 experiment~\citep{gando2020first, gando2021nylon,KamLANDIBD2023}.

\section{Data selection and background estimation} \label{sec:antiv}

\subsection{Electron antineutrino selection}
KamLAND uses the inverse beta decay (IBD) process to detect electron antineutrino interactions: $\bar{\nu}_e + p \rightarrow e^+ + n$. When this process occurs in the outer balloon of the detector, it produces a delayed coincidence (DC) signal that KamLAND can detect. The positron generated in the interaction will emit photons via scintillation before almost immediately annihilating with an electron, giving a prompt signal of two 511\,keV photons ($E_p$). The neutron produced in the interaction will capture on a proton (or carbon atom) with a mean capture time of 207.5\,$\pm$\,2.8\,$\mu$s, resulting in a delayed signal of a 2.22 (4.95)\,MeV photon~\citep{abe2010production}. This delayed coincidence pair is separated both in space ($\Delta$R\,$<$\,200\,cm) and time (0.5\,$\mu$s\,$<$\,$\Delta$T$\,<$\,1000\,$\mu$s). At low energies, the antineutrino energy can be approximated as the sum of the prompt energy (positron kinetic energy + annihilation energy = 2m$_e$) + 0.782\,MeV, ignoring the relatively small kinetic energy of the neutron~\citep{asakura2015study}. To search for neutrinos at 200 MeV, we use the energy cut 0.9\,MeV\,$<$\, $E_p$\,$<$\,200\,MeV to conservatively cover uncertainties in the kinetic energy of the positron, the kinetic energy of the neutron, and energy scale uncertainties even if $\langle$$E_p$$\rangle$ = 172 MeV.
The DC temporal and spatial separation, as well as the known energies of the resulting photons, greatly reduces background signal contamination and results in a high detection efficiency at low energies. 

To further isolate signal events, cuts are placed on the energy of the photon resulting from the neutron capture ($E_d$) such that 1.8 (4.4)\,MeV\,$<$\, $E_d$ $<$\,2.6 (5.6)\,MeV for capture on protons (carbon). A further cut requires the DC pair to be found within a 6\,m radius spherical volume in the center of the outer balloon. During the KamLAND-Zen experiment, an additional cut is used to avoid backgrounds from the inner balloon, removing a 2.5\,m-radial region around the vertical axis of the detector, extending from its top to its center, as well as a 2.5\,m-radius sphere around the center of the inner balloon. Finally, a likelihood-based selection is used to distinguish DC pairs from background events~\citep{gando2013reactor}. GRB221009A took place during the KamLAND-Zen 800 experiment, so the inner balloon cut (the purple region in Fig. \ref{fig::detector}) is included. The selection efficiency is shown in Fig.~\ref{fig::detector}(right) and converges to 77.4\% during this time, taking into account the loss in fiducial volume due to the inner balloon cut.



\section{Time-Coincident Event Search} \label{sec:candidates}
We perform a time-coincident IBD search looking for antineutrino events in the detector that arrive in various time windows surrounding the GRB trigger time $T_0$. These windows cover timing differences in theoretical production mechanisms for neutrinos compared to photons, as well as the neutrino time-of-flight delay, with different time windows allowing the analysis to cover a wider range of possible production mechanisms. For each time window, we determine the number of IBD events observed by KamLAND, and then create the Feldman-Cousins confidence belt at the 90\% confidence level~\citep{feldman1998unified} from the number of observed IBD events in the time window, as well as the background IBD event rate in the detector. To calculate the background rate, we look for IBD events over the whole of the KamLAND-Zen 800 period and divide that number by the detector livetime in that period. The expected number of observed IBD events can then be calculated for each time window. Model-independent antineutrino flux limits are then computed with the Green’s Function using $N_{90}$, the upper bound of the Feldman-Cousins confidence belt, for each time window:

\begin{equation}\label{eq::greenes}
\Psi_{90}(E_{\bar{\nu}_e}) = \frac{N_{90}} {N_T \sigma(E_{\bar{\nu}_e}) \epsilon_{live}(E_{\bar{\nu}_e})},
\end{equation}
where N$_{T}$ is the number of targets for IBD in the KamLAND detector, $\epsilon_{live}(E_{\bar{\nu}_e})$ is the detector efficiency, and $\sigma(E_{\bar{\nu}_e})$ is the IBD cross section~\citep{strumia2003precise}. Finally, the time-integrated antineutrino flux $F_{90}$ is calculated assuming various source neutrino emission models in a single bin from 1.8 to 200\,MeV:

\begin{equation}\label{eq::flux}
F_{90} = \frac{N_{90}} {N_T \int  \sigma(E_{\bar{\nu}_e}) \lambda(E_{\bar{\nu}_e})  \epsilon_{live}(E_{\bar{\nu}_e}) dE_{\bar{\nu}_e} },
\end{equation}
where $\lambda(E_{\bar{\nu}_e})$ is the normalized neutrino energy spectrum produced at the source. We choose power laws for this spectrum with indices from 1.5 to 3 in steps of 0.5.

\section{Results}
The conservative neutrino time-of-flight delay from GRB221009A, assuming a heaviest neutrino mass of 59 meV, and base-$\Lambda$CDM cosmological parameters from~\cite{Planck2018Results}, was calculated to be $\sim$30\,s. We selected 6 time windows, shown in Table \ref{table::timewindows}, that clearly account for the time-of-flight delay, and allow for a wide range of neutrino production mechanisms. We did not observe any neutrinos within any of the chosen time windows. With 0 observed neutrinos and a background rate we measured to be 2.23\,$\mu$Hz, the signal upper limits for each time window were calculated. 

\begin{figure}[H]
\centering
\includegraphics[width=0.65\textwidth]{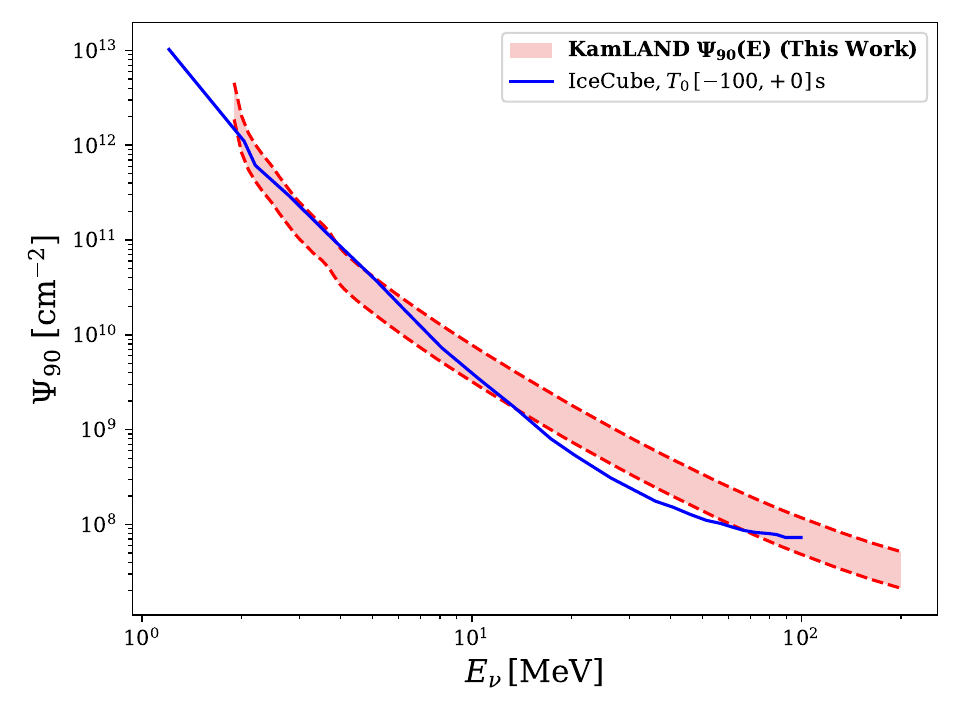}
\caption{The KamLAND 90\% confidence level model-independent Green's Function as a function of energy between 1.8 MeV and 200 MeV (red band) and IceCube ASTERIA 90\% confidence level upper limit (blue). The lower limit of the red band was determined from a time window of $T_0 [-1,\,+14]\,$days, while the upper limit of the band was calculated using a time window of $T_0 [+0,\,+327]\,$s. The KamLAND result is consistent with other non-detection KamLAND analyses, as in \cite{Abe2022}.}
\label{fig::greens}
\end{figure}

\begin{deluxetable*}{ccccc}[h]
    \color{\cc}
    \label{table::timewindows}
    \tablecaption{The different time windows used in the analysis. For each time window, the number of observed neutrinos, the signal upper limit, number of \nue we expect to see from the background rate, and the calculated flux upper limit (scaled by $E^2$ and using a power law index of 2) are shown.}
    \tablewidth{0pt}
    \tablehead{
      \colhead{Time Window}   & \colhead{\nue$_{obs}$} & \colhead{N$_{90}$} & \colhead{Expected \nue} & \colhead{$E^2 F_{90}(E)$ ($\gamma$ = 2)[GeV cm$^{-2}$]}
    }
    \startdata
    $T_0 [+0,+327]$ s  & 0 & 2.44 & 0.001 $\pm$ 0.027 & $3.12 \times 10^6$  \\ [0.5ex]
    $T_0 \pm 500$ s  & 0 & 2.43 & 0.002 $\pm$ 0.047 & $3.11 \times 10^6$ \\ [0.5ex]
    $T_0 [-200,+2000]$ s & 0 & 2.43 & 0.005 $\pm$ 0.070 & $3.11 \times 10^6$ \\ [0.5ex]
    $T_0 [-1,+2]$ hr & 0 & 2.41 & 0.024 $\pm$ 0.155 & $3.10 \times 10^6$ \\ [0.5ex]
    $T_0 \pm 1$ day & 0 & 2.05 & 0.381 $\pm$ 0.619 & $2.80 \times 10^6$ \\ [0.5ex]
    $T_0 [-1,+14]$ day & 0 & 1.00 & 2.861 $\pm$ 1.690 & $1.34 \times 10^6$  \\ [0.5ex]
    \enddata
 \color{black}
\end{deluxetable*}
The model-independent Green’s` Function is shown in Fig. \ref{fig::greens}, where the band represents the space between the lowest and highest limits between the different time windows. This is compared to the IceCube low-energy model-independent limit~\citep{kruiswijk2023icecube}.

\begin{figure}[H]
\centering
\includegraphics[width=0.8\textwidth]{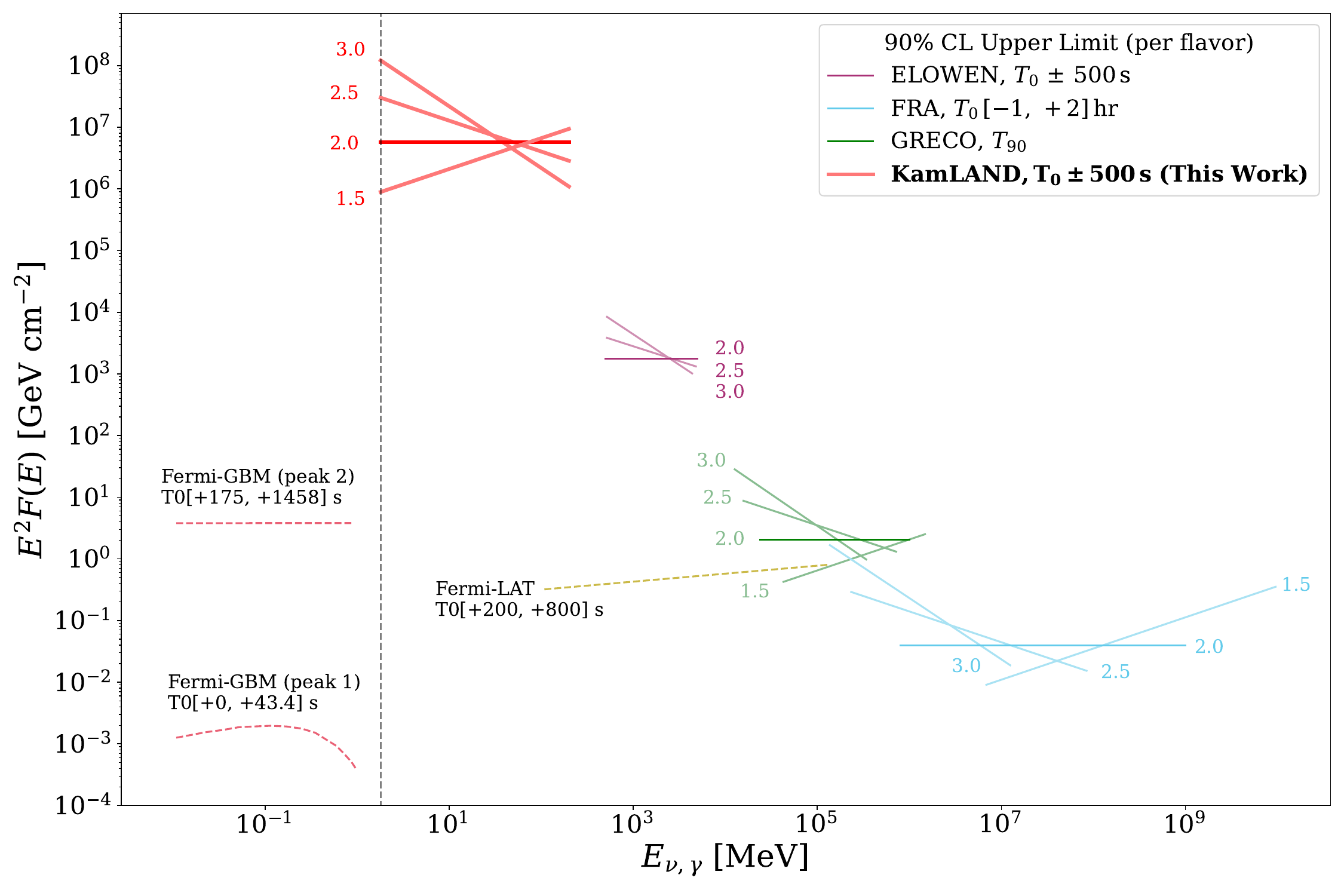}
\caption{The 90\% confidence level per-flavor neutrino flux from this work, shown in red, and IceCube analyses, shown in purple, green, and cyan \citep{IceCubeGRB}. The horizontal dashed lines are gamma-ray observations (not upper limits) from Fermi. The vertical dashed line shows the IBD lower limit at 1.8 MeV. The curve labels identify the source emission power-law spectra, corresponding to the power law index. The time windows for each curve are labeled. We chose $T_0 \pm 500\,$s to match IceCube's ELOWEN analysis, as this was the lowest-energy IceCube result for a power-law source emission spectrum.}
\label{fig::result}
\end{figure}

Finally, we calculated the electron antineutrino flux for multiple source emission power law spectra, choosing the time window to be $T_0$ $\pm$ 500\,s. The power law results are compared with IceCube upper limits, as well as gamma-ray observations from Fermi, in Fig. \ref{fig::result}. These results are consistent with IceCube's non-observation of neutrinos from the GRB~\citep{IceCubeGRB}.

\section{Discussion} \label{sec:discussion}

GRB221009A has favorable properties, such as a small jet opening angle pointing at Earth ~\citep{LHAASOScience}, so the lack of neutrino detection indicates that future events would need to be closer to Earth to stand a chance of being observed. Models for low-energy neutrino emission from GRBs exist for both isotropic and beamed emission, but the increase in flux from the beaming, without having been able to detect those neutrinos, means that it is unlikely to be possible to detect isotropic neutrino emission unless the event was extremely close to Earth. 

Although we did not observe any neutrinos from GRB221009A, this analysis provides a pipeline for detecting neutrinos from future GRB events, as well as providing a useful benchmark for comparison, due to the unique energetics of GRB221009A.

\section{Conclusion} \label{sec:conclusion}
We present a search for electron antineutrinos using the KamLAND detector between 1.8 and 200 MeV coincident with GRB221009A, the brightest gamma-ray burst ever observed. We search for events within multiple time windows surrounding the GRB trigger time, corresponding to multiple possible \nue production mechanisms, as well as accounting for the neutrino time-of-flight delay. No coincident antineutrinos are found, but we are able to set upper limits on the \nue flux from the GRB using multiple source emission spectra. These limits, as well as model-independent limits, are compared to limits placed by IceCube, and are found to be comparable.

\section{Acknowledgments}
\begin{acknowledgments}
The \mbox{KamLAND-Zen} experiment is supported by JSPS KAKENHI Grants No. 21000001, No. 26104002, No. 24H02237, and No. 19H05803; the U.S. National Science Foundation awards no. 2110720 and no. 2012964; the Heising-Simons Foundation; the Dutch Research Council (NWO); and under the U.S. Department of Energy (DOE) Grant No.\,DE-AC02-05CH11231, as well as other DOE and NSF grants to individual institutions. The Kamioka Mining and Smelting Company has provided service for activities in the mine. We acknowledge the support of NII for SINET4.
\end{acknowledgments}
\newpage

\bibliography{Main}{}
\bibliographystyle{aasjournal}

\end{document}